\documentclass[conference]{IEEEtran}
\IEEEoverridecommandlockouts
\usepackage{amsmath,amssymb,amsfonts}
\usepackage{algorithmic}
\usepackage{graphicx}
\usepackage{textcomp}
\usepackage[table]{xcolor}
\def\BibTeX{{\rm B\kern-.05em{\sc i\kern-.025em b}\kern-.08em
    T\kern-.1667em\lower.7ex\hbox{E}\kern-.125emX}}

\usepackage{amsthm}
\usepackage{comment}
\usepackage[backend=biber]{biblatex}

\newtheorem{lemma}{Lemma}
    
\begin{document}

\title{Empirical Analysis of GPU Frequency Behavior Under ML Workloads}

\author{
    \IEEEauthorblockN{Truong-Thanh Le\IEEEauthorrefmark{2}, 
                      Hoang-Loc La\IEEEauthorrefmark{4}, 
                     Amir Taherkordi\IEEEauthorrefmark{2},
                     Frank Eliassen\IEEEauthorrefmark{2},
                     Phuong Hoai Ha\IEEEauthorrefmark{4} and
                     Peiyuan Guan\IEEEauthorrefmark{2}} 

    \begin{tabular}[t]{@{}c@{}}
    \IEEEauthorblockA{\IEEEauthorrefmark{2} 
                      Department of Informatics \\
                      \textit{University of Oslo}\\
                      Oslo, Norway \\
                      \{truongl,amirhost,frank,peiyuang\}@ifi.uio.no} 
    \end{tabular}\nobreak\qquad

    \begin{tabular}[t]{@{}c@{}}
    \IEEEauthorblockA{\IEEEauthorrefmark{4} 
                      Department of Computer Science \\
                      \textit{UiT The Arctic University of Norway}\\
                      Tromsø, Norway \\
                      \{hoang.l.la,phuong.hoai.ha\}@uit.no}
                      \end{tabular}   
}

\maketitle

\begin{abstract}
This work presents ongoing research on the frequency-scaling behavior of NVIDIA GPUs when executing ML/AI workloads. Our preliminary findings show that, on lower‑performance GPUs, the operating frequency is strongly affected by the recent workload history---typically within an 80ms window. This behavior challenges a common assumption underlying several state‑of‑the‑art ML latency‑prediction techniques, which treat individual GPU kernel latencies as independent and therefore estimate total execution time by summing isolated per‑kernel measurements. Our results indicate that such an assumption does not always hold, as the GPU’s dynamic frequency scaling introduces inter‑kernel dependencies. We also outline several promising directions for leveraging this observation in future work, including improved latency‑prediction models, GPU kernel‑reordering strategies, and NAS‑driven guidelines for frequency/latency/energy‑aware model design.
\end{abstract}


\section{Introduction}
These days, modern GPUs span a wide range of performance levels, from compact mobile chips to large data‑center accelerators. 
Due to these differences in thermal design, power delivery, and efficiency requirements, the runtime behavior of such devices can vary significantly, leading to distinct performance patterns---especially in how they manage operating frequency under ML and AI workloads. Hence, understanding how the dynamic frequency‑scaling mechanisms behave on these different classes of GPUs is crucial for accurately characterizing their performance under ML and AI workloads.

When running ML/AI workloads, the GPU often scales its operating frequency under the control of an internal management core, which dynamically adapts the frequency based on recent or ongoing activity. This controller typically considers factors such as the number of arithmetic operations executed, the volume of data transferred across the DRAM–L1/L2 cache hierarchy, thermal condition, power limit, and a variety of hidden factors \cite{le2026pm2lat}. Unfortunately, understanding the precise decision-making policies of this controller is difficult, as both its firmware and the GPU‑kernel implementations of modern ML/AI models are closed‑source, leaving these mechanisms largely opaque to researchers.

Due to this complexity, most state‑of‑the‑art methods for predicting the performance of ML/AI workloads on modern GPUs simply ignore such behavior, assuming instead that each kernel operates independently. For example, 
Paleo \cite{qi2017paleo}, Habitat \cite{geoffrey2021habitat}, NeuSight \cite{lee2025forecasting}, and PM2Lat \cite{le2026pm2lat} all adopt a similar strategy for estimating model latency on specific devices: they categorize GPU kernels into multiple groups, assign a predictive model or parameter set to each category, and then estimate the total execution time by summing the predicted latency of individual kernels. Although several approaches attempt to capture end‑to‑end behavior by treating a model as a sequence of kernels and applying sequence‑based learning techniques—such as graph neural networks \cite{zhao2025perfseer}, or LSTMs \cite{assali2024multivariate}—to predict total latency, these methods still fall short of generalizing across different GPU architectures and model types, making them less efficient and limiting their ability to accurately capture real performance variations in practice.

Our ongoing work examines GPU frequency‑scaling behavior in greater detail under modern AI/ML workloads, particularly on lower‑performance or thermally constrained GPUs, where we observe that the operating frequency is heavily influenced by short‑term workload history rather than static hardware limits. Specifically, for laptop‑class GPUs (e.g., NVIDIA RTX 3060M) and thermally constrained accelerators (e.g., NVIDIA T4, L4), we find that the controller considers approximately an 80ms window of recent workload activity when selecting the frequency for the next 20ms interval. In other words, every 20ms the GPU adjusts its operating frequency based on the characteristics of the kernels executed during the preceding 80ms. 

\section{Experiments}\label{sec.Exp}
In this section, we conduct extensive experiments to analyze the frequency‑scaling behavior of several modern GPUs, including the NVIDIA RTX 3060M, T4, L4, A100, and RTX Pro 6000 Blackwell. Detailed specifications for each device are provided in Table \ref{tab:GPUInfo}. 

\begin{table}[ht]
    \centering
    \caption{Specification of tested GPUs.}
    \label{tab:GPUInfo}
    \small
    \begin{tabular}{|c|c|c|c|c|c|}
        \hline
         & \cellcolor{gray!32}3060M &  \cellcolor{gray!32}T4 &  \cellcolor{gray!32}L4& \cellcolor{gray!32}A100& \cellcolor{gray!32}6000\\ 
         \hline
         Max Freq (GHz)& 2.090&1.590& 2.040&1.410&2.617\\
         \hline
         FP32 (TFLOPs)& 16.05&8.141&30.29&19.49&126.0\\
         \hline
         DRAM BW (GB/s)& 336&320&300&1560&1790\\
         \hline
         MEM (GB)& 6& 16&24&40&96\\
         \hline
         L2 (MB)& 3& 4&48&40&48\\
         \hline
         SM Count & 30& 40&58&108&188\\
         \hline
         No.CUDA.Cores & 3840& 2560&7242&6912&24064\\ 
         \hline
         Power (W) & 130 & 70 & 70 & 400 & 600 \\
         \hline
    \end{tabular}\\
    6000: RTX PRO 6000 Blackwell
\end{table}

\subsection{Kernel-level Frequency/Throughput analysis}
As mentioned in the paper of PM2Lat \cite{le2026pm2lat} and NeuSight \cite{lee2025forecasting}, when increasing the size of a kernel---for example, a Matrix Multiplication (MatMul - $(M,K)\times(K,N)=(M,N)$) kernel---by launching more waves or enlarging the hidden size, the achieved throughput typically increases following a rational‑function‑like trend. In this subsection, we extend this analysis by also incorporating the GPU’s operating frequency, providing a more complete view of how both throughput and frequency scale with growing workload size. 
For each data point, we sample at least 500ms of execution time, with at least 25 times of measurement, to ensure that the collected results reflect stable and representative behavior.

\begin{figure}
    \centering
    \includegraphics[width=\linewidth]{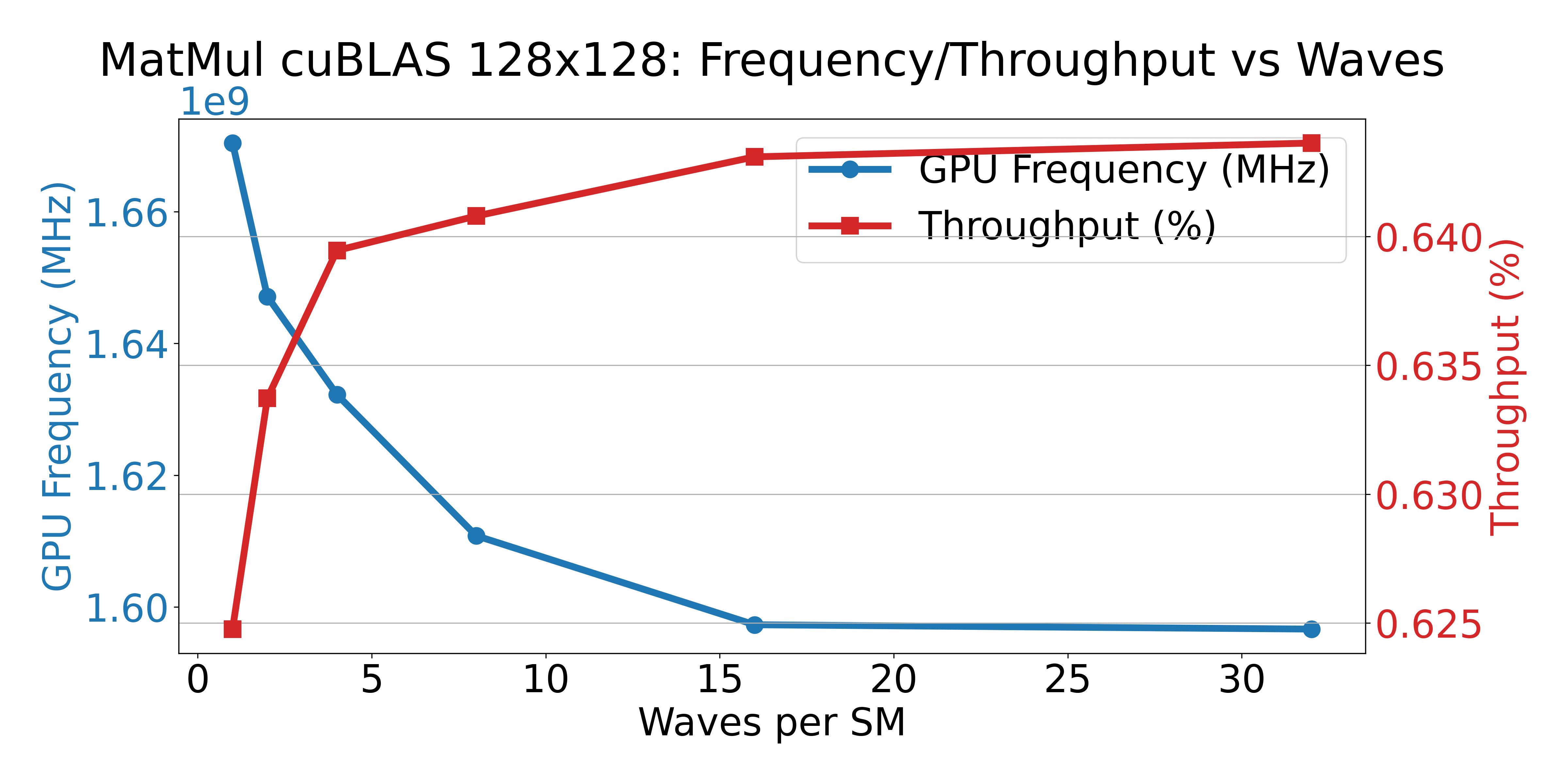}
    \caption{Frequency/Throughput vs Wave (RTX3060M).}
    \label{fig:RTX3060M.S1.1}
\end{figure}

\begin{figure}
    \centering
    \includegraphics[width=\linewidth]{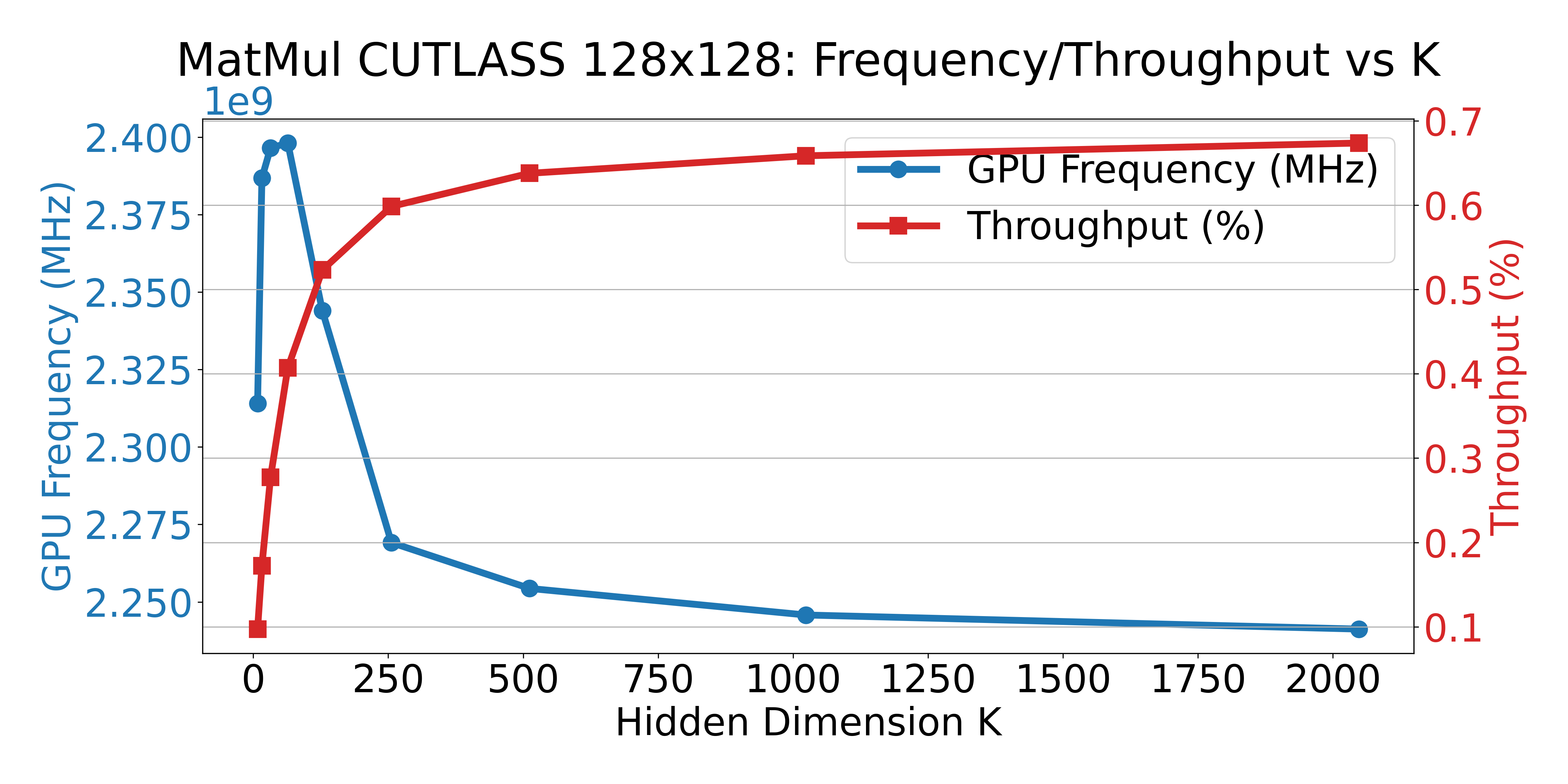}
    \caption{Frequency/Throughput vs Hidden Dimension (RTX PRO 6000 Blackwell).}
    \label{fig:6000.S1.2}
\end{figure}

As we can see in Figures \ref{fig:RTX3060M.S1.1} 
and \ref{fig:6000.S1.2}, the throughput of the kernel on both 3060M and 6000 Blackwell increases as we scale the hidden dimension or the number of waves launched per Streaming Multiprocessor (SM), following a rational‑function‑like trend. The operating frequency exhibits a similar rational pattern but with a downward trajectory, decreasing gradually as the workload becomes heavier. We experienced that the same frequency/throughput behavior applies to T4, L4 and A100.

\begin{lemma}
    When increasing the workload of a GPU kernel, the achieved throughput is increasing while the frequency is decreasing, following a rational trend.
\end{lemma}

\subsection{Frequency analysis with identical computation workload but different implementation ands configurations}

Even when two kernels perform the same computational workload, their achieved throughput and operating frequency can differ significantly. Prior work (e.g., PM2Lat \cite{le2026pm2lat}) has shown that kernel implementations differ in memory‑access patterns, tiling strategies, and microarchitectural behavior, which leads to measurable throughput variation. We extend this observation by showing that these implementation differences also produce distinct operating frequencies. Using a fixed MatMul workload $(2048,2048)\times(2048,2048)=(2048,2048)$, we compare kernels generated by cuBLAS and CUTLASS across several tile configurations. As shown in Figure 3, different configurations yield different steady‑state frequencies on the NVIDIA L4, despite having identical FLOP counts. This demonstrates that the GPU’s frequency controller responds not only to computational intensity but also to the kernel’s memory behavior and implementation‑specific characteristics.

\begin{lemma}
    For kernels with identical computational workloads, different implementations or configurations can produce different operating frequencies due to variations in memory‑access patterns.
\end{lemma}

\begin{figure}
    \centering
    \includegraphics[width=\linewidth]{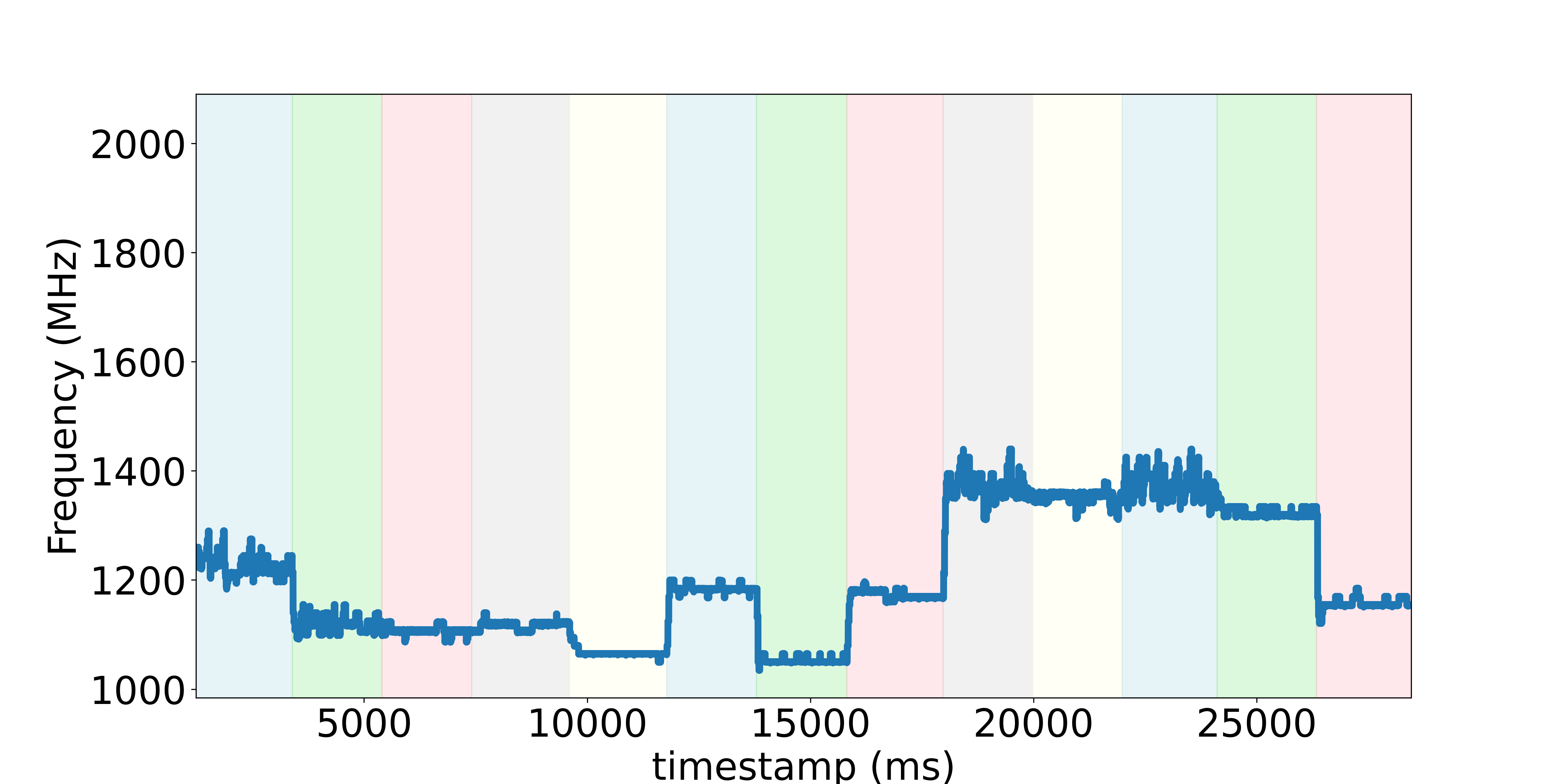}
    \caption{Frequency analysis on different kernel configurations (L4).}
    \label{fig:L4.S3.1}
\end{figure}

\subsection{Window-sized Frequency-Scaling Decision}
In the previous subsection, we analyzed how the GPU selects operating frequencies for different workloads in an isolated manner. In this subsection, we investigate how the frequency changes in response to abrupt shifts in workload characteristics—for example, when transitioning from a compute‑intensive kernel to a memory‑intensive one. To examine this behavior, we repeatedly execute a MatMul kernel for several iterations and then repeatedly execute a ReLU kernel, allowing us to observe how the GPU adjusts its operating frequency as the workload pattern changes.

\begin{figure}
    \centering
    \includegraphics[width=\linewidth]{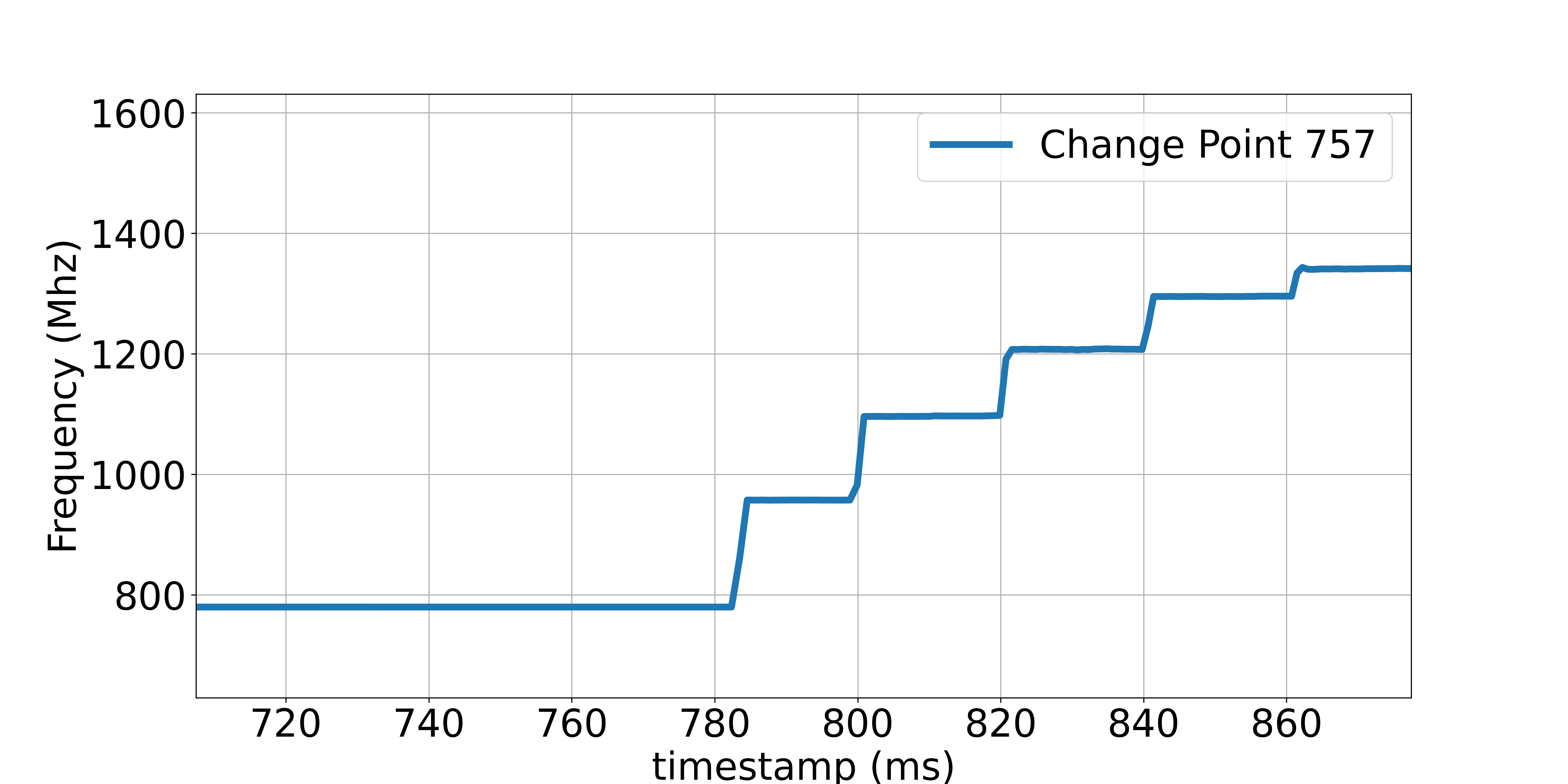}
    \caption{Frequency-scaling corresponding to the change in workload (T4).}
    \label{fig:T4.S2.1}
\end{figure}

\begin{figure}
    \centering
    \includegraphics[width=\linewidth]{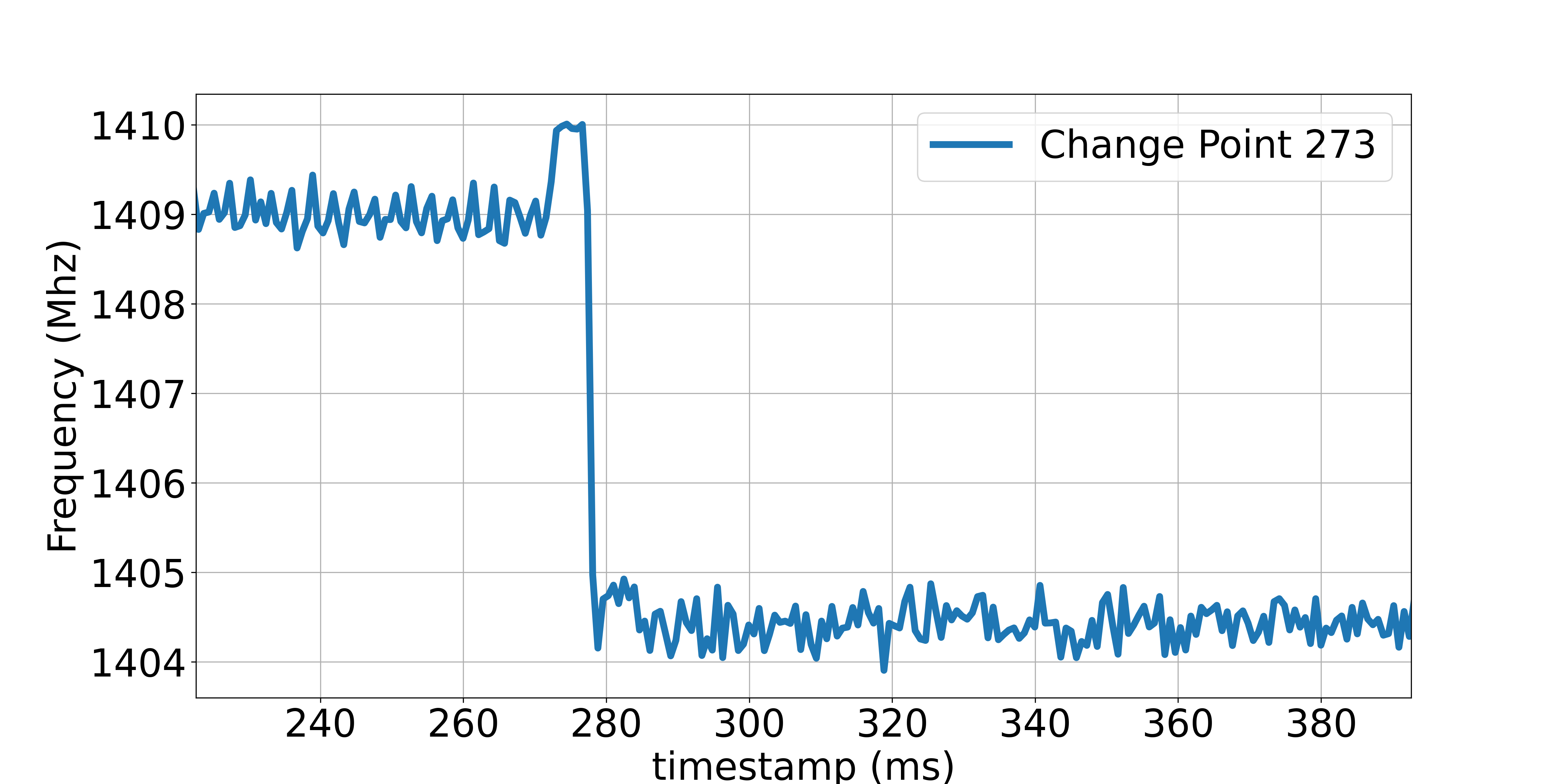}
    \caption{Frequency-scaling corresponding to the change in workload (A100).}
    \label{fig:A100.S2.1}
\end{figure}

Figures \ref{fig:T4.S2.1} and \ref{fig:A100.S2.1} show the results of this experiment, where frequency is measured in MHz and timestamps in milliseconds. As we can see, after the workload transition, the NVIDIA T4 continues operating at approximately 800MHz for several milliseconds before gradually increasing to more than 1300MHz, which corresponds to the operating frequency of the ReLU kernel. This delayed adjustment behavior also appears on the RTX 3060M and the L4. In contrast, on the A100 and the RTX PRO 6000 Blackwell, the operating frequency shifts immediately when the workload changes, reflecting their higher thermal headroom and more responsive frequency‑scaling mechanisms.

From additional experiments conducted on the RTX3060M, T4, and L4, we observe that the operating frequency on these GPUs changes at fixed 20ms intervals. Furthermore, the frequency stabilizes at the characteristic operating frequency of the ReLU kernel approximately 80ms after the first frequency‑transition point. Importantly, during this 80ms interval, only ReLU kernels are executed, indicating that the frequency‑scaling decision depends on the recent 80ms window of workload history rather than the immediate kernel. 

\begin{lemma}
    On lower‑performance GPUs and thermal- constraint GPUs, the frequency is updated every 20ms, and the controller’s effective decision‑making window spans 80ms.
\end{lemma}

It is important to note that when we repeatedly run a kernel, the GPU frequency stabilizes at a characteristic value with only minor fluctuations; we refer to this value as the kernel’s operating frequency ($f$). Hence, we assume that if there exists a set of descriptive information $I$ about the kernel, then its operating frequency can be modeled as
\begin{equation}
    f=F(I)
\end{equation}
where $F$ represents the (unknown) decision function implemented by the GPU’s internal frequency‑scaling controller. We already know that the amount of computation and the volume of data transfer are major components of $I$. However, relying solely on these two parameters is insufficient to approximate $F$, as additional hidden factors---such as memory‑access patterns across the DRAM–L1/L2 hierarchy, specific kernel implementations, tiling strategies, and other microarchitectural behaviors---also influence the resulting operating frequency.

With an 80ms window, the controller observes multiple kernels, each with its own operating frequency, e.g. $f_1,f_2,...,f_n$. This raises the central question: \textit{how can we predict the GPU’s frequency for the next 20ms interval?} To investigate this, we design an experiment in which we repeat a sequence of kernels whose individual operating frequencies are already known. By repeating the same sequence, each 80ms window contains an identical set of kernels, allowing the GPU to reach a stable and consistent frequency‑scaling pattern across windows. This controlled setup enables us to precisely measure how the controller combines the operating frequencies of individual kernels to determine the frequency for the next 20ms segment. For example, when running on the RTX 3060M, a combination of kernel A with $f_A=$1580MHz and kernel B with $f_B=$1680MHz results in a stable GPU frequency of approximately 1615MHz. Notably, within each 80ms window, the average execution time of kernel A ($t_A$) and kernel B ($t_B$ is about 60ms and 20ms, respectively. Hence, if we assume that the controller aggregates the operating frequencies of the kernels appearing in the window, a natural hypothesis is that the resulting frequency is determined by a latency‑weighted average of the individual kernel frequencies. Under this assumption, the expected operating frequency would be:
\begin{equation}
    f_{pred}=\frac{t_A*f_A+t_B*f_B}{t_A+t_B}=\frac{60*1580+20*1680}{60+20}=1605
\end{equation}
Then, the prediction error is only about 10MHz, corresponding to an error rate of approximately 0.6\%. This characteristic also holds when increasing the number of kernels in the sequence—the same latency‑weighted behavior persists. Moreover, when evaluating other low‑performance or thermally constrained GPUs, we observe the same pattern, indicating that this frequency‑scaling mechanism is consistent across devices in this category. Additionally, we also find that memory‑intensive kernels consistently exhibit a characteristic operating frequency of the GPU, suggesting that the controller prioritizes maximum frequency when the workload is bottlenecked by memory bandwidth rather than computation.

\begin{lemma}
    On lower‑performance GPUs and thermally constrained GPUs, if the operating frequency and latency of every kernel within an 80ms window are known, the GPU’s frequency for the next 20ms interval can be estimated using a weighted formula based on the proportion of time each kernel occupies in the window.
\end{lemma}
    
\begin{lemma}
    The operating frequency of the memory-intensive kernels are always the highest achievable frequency of the GPU.
\end{lemma}

However, another problem arises: \textit{how can we measure the operating frequency of an individual kernel?} This would be straightforward if we could isolate the kernel and run it repeatedly until the GPU stabilizes at its characteristic operating frequency. Unfortunately, in modern AI workloads this isolation is rarely possible. When invoking a single layer, multiple kernels are typically launched in sequence—for example, a convolutional layer may execute at least three kernels---making it infeasible to extract a clean, standalone frequency value for any single kernel.

\section{Discussions: Potential application}
After performing extensive experiments in Section \ref{sec.Exp}, we conclude that the operating frequency of the GPU is not determined solely by the characteristics of the current kernel, but rather by a combination of the recent workload history and the internal scheduling logic of the frequency‑scaling controller. This directly contradicts the assumption adopted by current state‑of‑the‑art latency‑prediction methods, which treat the execution time of all kernels as independent in all GPUs. Consequently, our findings suggest a promising new direction: instead of assuming kernel‑level independence, one can first estimate the GPU’s operating frequency for each 20ms interval---based on the 80ms workload history---and then predict per-kernel latency using this dynamic frequency estimation. In addition, the operating frequency of the GPU also affects its power consumption. Hence, by estimating the GPU frequency for each 20ms interval, we can further extend this approach to predict the workload's power consumption.

Moreover, the analysis also opens a potential direction for task scheduling on GPUs. Specifically, when multiple parallel, independent tasks are submitted to the GPU,
the scheduler can reorder the kernels to favor higher running frequencies during periods dominated by compute‑intensive kernels, thereby reducing their latency, while allowing lower frequencies during periods dominated by memory‑intensive kernels, where performance is less sensitive to frequency. This frequency‑aware scheduling strategy has the potential to improve both performance and energy efficiency for mixed workloads. 

Finally, instead of relying solely on the available ML workloads to optimize latency and energy consumption, we consider a Neural Architecture Search (NAS)–based approach that leverages our findings. In particular, when using NAS to deploy ML models on low‑performance or thermally constrained GPUs, the search process can incorporate our frequency‑scaling observations to guide model selection. For example, NAS can favor architectures whose kernel sequences produce more favorable frequency patterns---such as longer compute‑intensive segments that benefit from higher operating frequencies or balanced workloads that avoid frequent down‑scaling events. By integrating dynamic frequency behavior into the objective function, NAS can more effectively explore architectures that optimize not only accuracy and FLOPs but also runtime, energy efficiency, and stability under real GPU operating conditions.

\section{Conclusion}
In this work, we conducted an extensive empirical study of GPU frequency‑scaling behavior under modern ML/AI workloads. Our results reveal that lower‑performance and thermally constrained GPUs adjust their operating frequency every 20ms based on an 80ms history window, introducing inherent inter‑kernel dependencies that contradict the kernel‑independence assumption used by current latency‑prediction methods. These findings open several promising directions, including improved frequency‑aware latency and power prediction, kernel‑reordering strategies for mixed workloads, and NAS‑guided model design for devices with constrained thermal or performance budgets. Overall, our study highlights the need to account for dynamic frequency behavior when modeling, optimizing, or deploying ML workloads on modern GPUs.

\section*{Acknowledgment}
This work was supported in part by the Norwegian Research Council under Grant 322473 (AirQMan project) and the European Commission under Grant 101086541 (MISO project).

\printbibliography

\end{document}